Getting a head start: the slime mold, *Physarum polycephalum*, tune foraging decision to motivational asymmetry when faced with competition

Stirrup E.[1] & Lusseau D.[1,*]

[1]Institute of Biological and Environmental Sciences, University of Aberdeen, Aberdeen AB24 2TZ, UK.
[*]Corresponding author: d.lusseau@abdn.ac.uk

**Abstract**

Slime mould plasmodia can adjust their behaviour in response to chemical trails left by themselves and other *Physarum* plasmodia. This simple feedback process increases their foraging efficiency. We still do not know whether other factors influence plasmodium behaviour in realistic competition settings. Here we designed a competition experiment where two plasmodia had to find one food source in a common environment. As previously shown, the time it took plasmodia to find food depended on their hunger motivation. However, the time it took a plasmodium to start looking for food depended on its motivation and the motivation of its competitor. Plasmodia always initiated foraging quicker if they were in the presence of a competitor and the quickest if they were hungry and in the presence of a satiated competitor. The time it took to arrive to the food was not influenced by whether they were alone or with a competitor. Ultimately, this complex competition response benefited the hungry plasmodia as they had a 4:1 chance of finding the food first. The sensory ecology of *Physarum polycephalum* is more complex than previously thought and yields complex behaviour in a simple organism.

Keywords: decision, foraging, hunger, *Physarum polycephalum*

## 1. Introduction

The presence of conspecifics can help individuals navigate their environment to maximise their fitness; simply through local enhancement [1] or social facilitation [2, 3]. The propensity to engage in either competitive or cooperative behaviours between conspecifics can be greatly influenced by individual hunger level as a motivational state [4, 5]. Inter-individual motivational asymmetry forces a dynamic trade-off between the benefits of foraging as a group and competition within the group for a food source [6]. Animals will take greater foraging risks when hungry [7] and chose to spend less time in larger groups after food deprivation [4]. Until recently, this has been widely studied in vertebrate species and thus, suggesting that neurological complexity supports the ability to make these rather complex decisions.

Seemingly simple organisms, that do not possess a centralised neural system (CNS), represent the majority of the Earth's taxa. There is increasing evidence for complex social behaviours in bacteria and microbes including cooperation [8] and competition [9]. The plasmodia of the acellular slime mould *Physarum polycephalum* (Physarum thereafter*)* is also capable of making complex foraging decisions based on trade-offs between risks, hunger level and food patch quality [10-13]. This body of work is redefining our fundamental understanding of the role of cognition in the emergence of complex social behaviour [14].

Physarum secretes a trail of slime following movement, which acts as an extracellular spatial memory [15]. This increases foraging efficiency as Physarum then avoids previously explored areas and conspecifics [15, 16]. We do not know how this competitive behaviour is integrated in Physarum decision-making processes. Competing individuals should modify their behaviour in accordance to the value of a resource and the resource holding potential (RHP) depending on asymmetries with competitors [17]. One of these asymmetries between individuals is hunger. If contestants are able to mutually gather information on RHP, then individual with the lower RHP would not engage in competition for which the resource reward would not be sufficient [18].

Previous studies on Physarum have shown that manipulating the hunger level of plasmodia influences foraging decisions based on trade-offs between motivation state, risk and food patch quality [10]. Here, we aimed to determine if foraging behaviour was influenced by the presence of another plasmodium with either the same motivation or not (hungry v satiated), when presented with only one food patch. We expected the pairing of Physarum in the testing arena to influence the time it took to find food as more extracellular slime trails would be produced. We also tested whether conspecifics presence affected other foraging decisions such as foraging initiation in accordance with RHP assessment strategies.

## 2. Materials and Methods

Sclerotia used to initiate the original culture of plasmodia were sourced from Carolina Biological Supply Company (Burlington, North Carolina, USA). Cultures were reared on 4% oatmeal agar food disks (4g oatmeal:1g agar:100ml water) in dark trays at room temperature (20-24°C). To test for influences of conspecific presence on behaviour, plasmodia were subject to either 'coupled' or 'alone' (control) treatments. Coupled treatments contained two plasmodia positioned 2cm from each other and 2cm from the food disk. Plasmodia could be either hungry or satiated and food disks were 8% (HFC) or 4% (LFC) oatmeal. Plasmodia were obtained by cutting 1cm$^2$ samples from the search front of the culture. 'Hungry' plasmodia were relocated to 1% agar plates for 24H prior to experimental run. 'Satiated' plasmodia were relocated to 8% oatmeal:1% agar plates for 24H before experimental runs. We ran 70 experiments, including 5 replicates for each treatment level, with a randomised running order (Table S1).

Plates, 8 at a time (Table S1), were placed in dark and moist conditions and left for 72H. We used a Logitech 1080p webcam suspended 20cm over the plates and a LED light board to capture still photographs of the plates every 15 minutes (the LED board was switched on automatically for 10sec. to do so). These photos were subsequently used to measure the distance (in pixels, using `imaqtool` in Matlab, then converted to cm) between plasmodia (distance between search fronts) and between plasmodia and the food disks (distance between the search front tip and the food disk edge). See Supplementary Methods for further details.

We determined three features of plasmodium behaviour: Start Time, Arrival Time and the Minimum Distance between plasmodia. Start Time (minutes) was the time at which plasmodium movement was first observed. Arrival Time (minutes) was the time at which the plasmodium first made contact with the food disk. Subsequent to this, the entire plasmodium biomass would move onto and engulf the food disk. No plasmodia were observed reaching the food source to then move away; some plasmodia never reached the food disk (NA – No Arrival). Minimum Distance was the smallest distance recorded between plasmodia during a run.

We used linear models to determine the effects of conspecifics on arrival time (log-transformed). If a plasmodium can detect and assess RHP of a conspecific, then conspecific presence will affect the start time of a given plasmodia and this, in turn, will be influenced by the given motivation of a neighbour. Thus, linear models were fitted to determine the effect of conspecific presence and motivation on start time (log-transformed). We finally determine whether motivation and motivation pairing (similar or mismatched) affected the minimum distance recorded between two conspecific plasmodia.

## 3. Results

When considering all treatment levels, hungry Plasmodia initiated foraging earlier and Plasmodia in a competitive setting always started foraging faster (Table S2-S3, Figure 1a). When considering coupled Plasmodia, start time depended on the motivation of the competitors as expected from RHP (Table S4). Hungry plasmodia with satiated neighbours started on average 18min after the start of the experiment. Hungry plasmodia with hungry neighbours took 27min, satiated plasmodia with satiated neighbours took 73min, and satiated plasmodia with hungry neighbours hungry took 256min on average (Figure 1b).

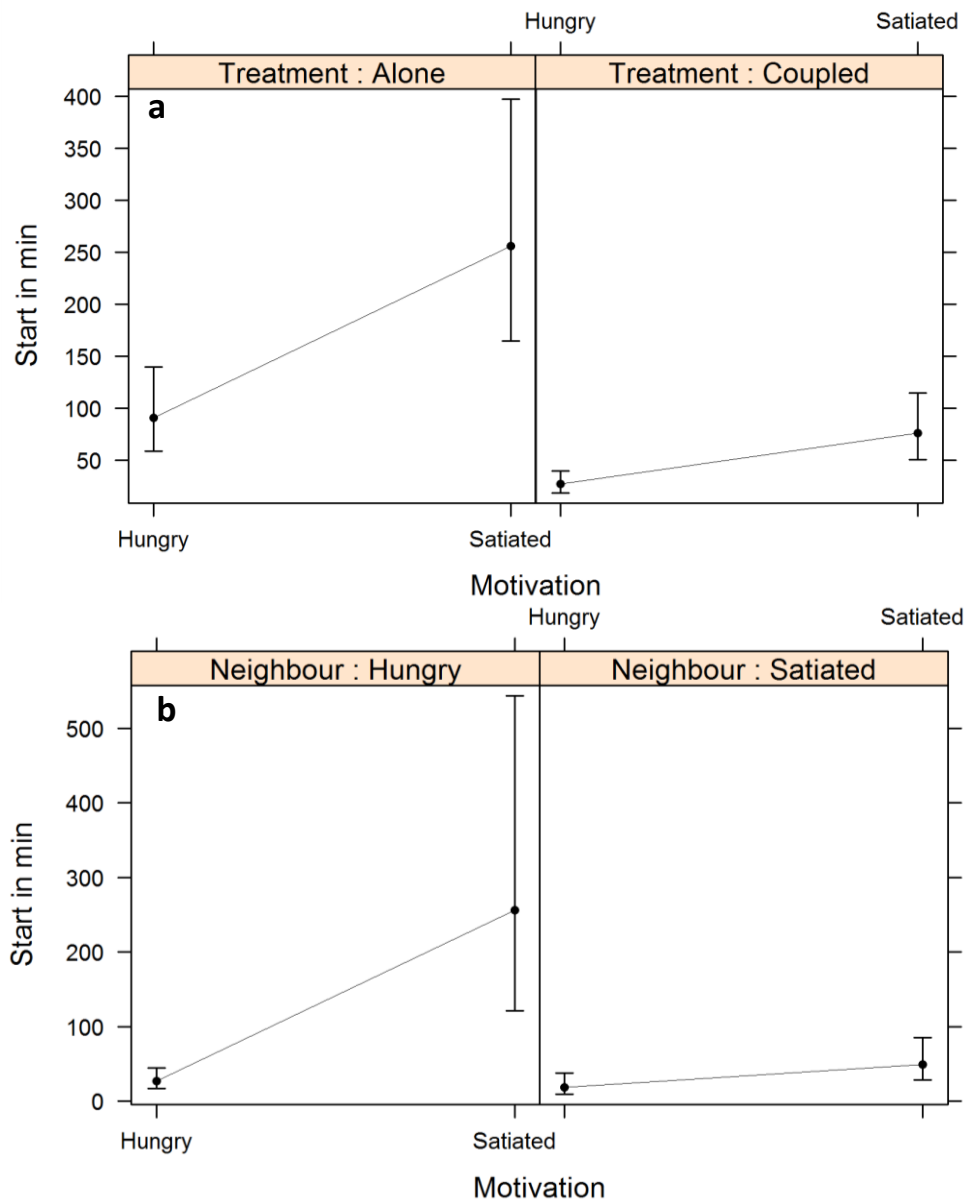

Figure 1. Best fitting model to explain (a) foraging initiation: back-transformed predicted Start time, with 95% confidence intervals for each level, depending in treatment level ($F_{1,89}$= 24.9, p<0.0001) and Physarum motivation ($F_{1,89}$= 20.1, p<0.0001) and (b) foraging initiation in a competitive setting: back-transformed predicted Start time, with 95% confidence intervals for each level, depending in the competitor's motivation (neighbour; $F_{1,50}$= 9.6, p=0.003), Physarum motivation ($F_{1,50}$= 19.0, p<0.0001), and their interaction ($F_{1,50}$= 4.0,

p=0.05).

Once they started foraging, no factor influenced the time it took plasmodia to find food (Table S5). However, in a competitive setting, Plasmodia with no asymmetry in motivation found the food faster than those that had motivational asymmetry when the resource was more lucrative (Table S6, HFC, Figure S1).

Hungry plasmodia were more likely to arrive first and more likely to do so if their neighbour was satiated (Table S7-S8, Figure 2a). In 10 instances, one plasmodium did not find the food disk within 72H. This was most likely to happen when the plasmodium that had first arrived to the food disk started as satiated and the food concentration was lower (LFC, Table S9, Figure 2b).

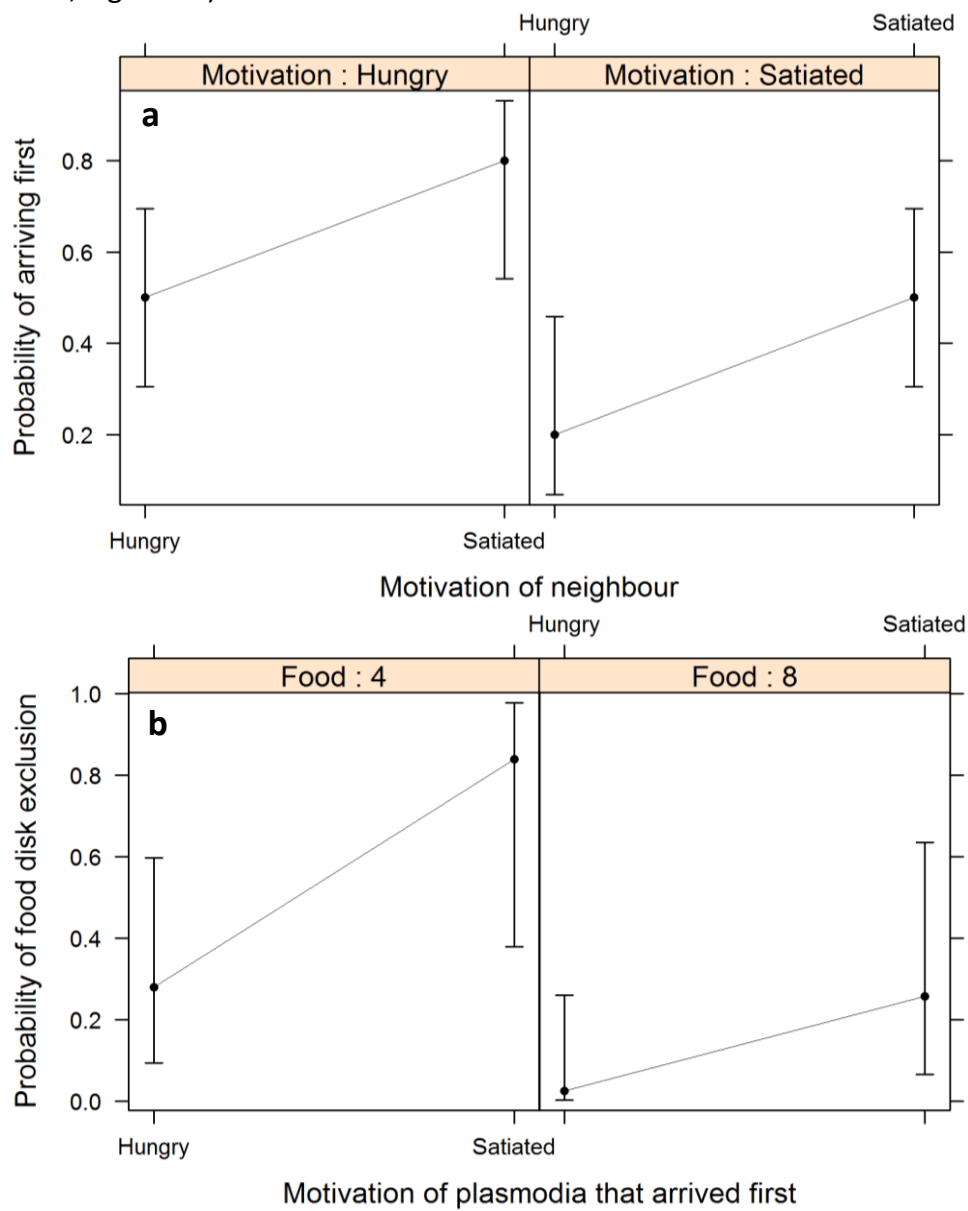

Figure 2. Best fitting model to explain (a) the probability that a plasmodium would be first to discover the food in a competitive setting, with 95% confidence intervals for each level, depending in its motivation ($\chi_1^2$= 75.5, p=0.02) and the motivation of its neighbour ($\chi_1^2$= 75.5, p=0.02) and (b) the probability that a plasmodium would be excluded from the food disk depending on the motivation of the plasmodium that arrived first ($\chi_1^2$= 26.1, p=0.01) and food concentration (LFC, 4g, and HFC, 8g, $\chi_1^2$= 32.5, p=0.04).

Plasmodia with motivation asymmetry stayed further apart during their foraging trip than the others (Table S10, Figure S2).

## 4. Discussion

Hungry plasmodia started foraging earlier and that led them to arrive earlier to the food disk. Unexpectedly, a plasmodium placed in a competitive setting started foraging faster than if it was alone. This is the first time that such behavioural modulation is observed outside taxa with a CNS. Importantly, this response to the presence of competitor is taking place without using the main known sensory mode for Physarum (extracellular slime). In light of this, it can be inferred that plasmodia have other means of conspecific detection than extracellular slime trails [15, 16]. Furthermore, plasmodia modulated their foraging decision based on not only the presence of a competitor, but also the motivation of that competitor. Being able to assess motivational asymmetry with conspecifics minimises the risks of energy loss in a competitive setting. This foraging initiation decision conferred an advantage to hungry plasmodia which were more likely to arrive first when set against a satiated plasmodium.

In most instances, plasmodia placed in the same arena maintained a minimum distance between their search fronts which was dependent on their motivational asymmetry. Similar studies in fish, showed that hungry individuals spent less time with a group of satiated individuals and strayed further from other individuals in general [4]. An asymmetry in motivation led here to more constraint in movement as search fronts maintained a greater distance from one another in a contained environment. This benefited hungry plasmodia which were then more likely to find food first.

Contrary to our predictions, food concentration did not affect foraging behaviour until a plasmodium had found the food disk. This suggests that plasmodia needed to make contact with a food source to analyse its nutritional value. Once they had found the disk, plasmodia that had started satiated were more like to be able to exclude the competitor from the food disk in low food concentration situations.

These results further compliment recent studies that show that plasmodia demonstrate flexibility in foraging strategies and complexity of decisions based on simple rules of thumb using both internal and external cues [10-12, 16]. The observed behavioural changes require Physarum to acquire information from its competitor before foraging, and any movement, was initiated [19] without using the usual approach to sense others [15]. The sensory ecology of Physarum is therefore richer than we previously thought. Physarum is forcing us to rethink the mechanisms by which living species make decisions [20]. This study breaks another barrier for non-CNS species, showing that motivational asymmetry can be appraised and that information used by simple species to make efficient foraging decisions.

**Funding Statement.** DL was supported in part by Scottish Funding Council grant HR09011 to the Marine Alliance for Science and Technology for Scotland.